\documentclass[12pt]{elsart}

\usepackage{amsmath}
\usepackage{amsfonts}
\usepackage{graphicx}

\begin{document}

\begin{frontmatter}

\title{Calculation of statistical entropic measures \\ in a model of solids}

\author[jsr]{Jaime Sa\~{n}udo}
\ead{jsr@unex.es} and
\author[rlr]{Ricardo L\'{o}pez-Ruiz}
\ead{rilopez@unizar.es}

\address[jsr]{
Departamento de F\'isica, Facultad de Ciencias, \\
Universidad de Extremadura, E-06071 Badajoz, Spain, \\
and BIFI, Universidad de Zaragoza, E-50009 Zaragoza, Spain}

\address[rlr]{
DIIS and BIFI, Facultad de Ciencias, \\
Universidad de Zaragoza, E-50009 Zaragoza, Spain}


\begin{abstract}
In this work, a one-dimensional model of crystalline solids based on 
the Dirac comb limit of the Kr\"onig-Penney model is considered. 
From the wave functions of the valence electrons, we calculate
a statistical measure of complexity and the Fisher-Shannon information 
for the lower energy electronic bands appearing in the system. All these magnitudes present
an extremal value for the case of solids having half-filled bands, a configuration where
in general a high conductivity is attained in real solids, such as it happens with
the monovalent metals.
\end{abstract}


\end{frontmatter}

\maketitle

The application of information theory measures to quantum systems is a subject of interest 
in the last years \cite{gadre1987,panos2005,sanudo2008-}.
Some relevant properties of the hierarchical organization of atoms \cite{panos2009,sanudo2009}
and nuclei \cite{lopezruiz2010} are revealed when these indicators are calculated on these many-body 
systems. On the one hand, they display an increasing trend 
with the number of particles, electrons or nucleons. 
On the other hand, they take extremal values on the closure of shells.
Moreover, in the case of nuclei, the trace of magic numbers is displayed 
by these entropic products such as the Fisher-Shannon information \cite{romera2004}
and a statistical complexity measure \cite{lopez1995}.
Also these statistical quantifiers have revealed a connection with physical measures, 
such as the ionization potential and the static dipole polarizability \cite{sen2007} in
atomic physics. All of them, theoretical and physical magnitudes, 
are capable of unveiling the shell structure of atoms, specifically the closure of shells in 
the noble gases.

A strategy to calculate these indicators is to quantify the discrete hierarchical organization 
of these multiparticle systems through the fractional occupation probabilities. 
These probabilities capture the filling of the shell structure of these systems.
From them, the different statistical magnitudes are derived. The metallic clusters is another 
system that has also been studied with this method \cite{sanudo2011}.
As in the case of atoms and nuclei, the shell structure of the valence electrons is well 
displayed by the spiky behavior of the statistical complexity and the magic numbers 
are unveiled by relevant peaks of the Fisher-Shannon information.

A different strategy to compute these entropic magnitudes is to use the probability density of 
the quantum system as the basic ingredient. This can be analytically obtained in some cases such 
as the H-atom \cite{sanudo2008-} or numerically derived in other cases from a Hartree-Fock scheme
\cite{borgoo2007,panos2007} or a density functional-theory for atoms and molecules \cite{parr1989}. 

In this work, we address the problem to calculate these statistical indicators in a solid by this
last strategy. For this purpose, the band structure of the solid has to be determined.
The Kr\"onig-Penney (KP) model \cite{kp1931} is a one-dimensional model of crystalline solids that 
presents a band structure sharing many properties with band structures of more sophisticated models.
Moreover, it also has the advantage that allows to analytically find such electronic band structure.

The KP model considers that electrons move in an infinite one-dimensional crystal where the positive 
ions are located at positions $x= na/2$ with $n=\pm 1,\pm 2,\ldots$, generating a periodic potential
of period $a$. A simplified version of the KP model is obtained when this periodic potential 
is taken with the form of the Dirac comb \cite{flugge1994}:
\begin{equation}
V(x)=\frac{\hbar^2}{m}\;\Omega\sum_{n=-\infty}^{+\infty}\delta(x+na),
\label{eq-Dirac}
\end{equation}
where $\hbar$ is the Planck constant, $m$ is the electronic mass and $\Omega$ is the
intensity of the potential. In this case,
the spatial part $\Psi(x)$ of the electronic wave function is determined from the time 
independent Schr\"odinger equation:
\begin{equation}
\left[-\frac{\hbar^2}{2m}\,\frac{d^2}{d x^2}\,+\,V(x)\right]\Psi(x) = E\Psi(x),
\label{eqSchro}
\end{equation}
where the eigenvalue $E$ is the energy of the eigenstate $\Psi(x)$.

For a periodic potential, the Bloch's theorem \cite{kittel1996} establishes the form of the general 
solution of the Eq. (\ref{eqSchro}). This is a plane wave, with wave number $K$, modulated by 
a periodic function, $u_K(x)$:
\begin{equation}
\Psi(x)=e^{iKx}u_K(x),
\label{eq-u_k}
\end{equation}
where $u_K(x)$ has the periodicity of the crystal lattice: $u_K(x)=u_K(x+a)$.
It implies the following translation property, 
\begin{equation}
\Psi(x+a)=e^{iKa}\Psi(x).
\label{eq-trans}
\end{equation}
Let us observe that we have the case of free electrons when $V(x)=0$
and the solutions of type (\ref{eq-u_k}) recover the plane wave form with a total 
wave number $k=K$, with $u_K(x)=constant$. It suggests that 
the solution of Eq. (\ref{eqSchro}) in the region $0 < x <a$, where $V(x)=0$, can 
be associated in some way with a wave number $k$ and then written in the general form:
\begin{equation}
\Psi(x)= A\sin kx+B\cos kx,
\label{eq-psi1}
\end{equation}
and, by the translation property (\ref{eq-trans}), this solution in the region $a < x < 2a$ is 
\begin{equation}
\Psi(x)= e^{iKa}[A\sin k(x-a)+B\cos k(x-a)],
\label{eq-psi2}
\end{equation}
with $A$, $B$ complex constants and $k=\sqrt{2mE/\hbar^2}$. 

Two boundary conditions must be fulfilled by $\Psi$ at the point $x=a$:
on the one hand, the continuity of the wave function and, on the other hand, 
the jump in the derivative provoked by the Delta function (\ref{eq-Dirac}). 
This gives the relations:
\begin{eqnarray}
\hskip 3cm \Psi(a+0) & = & \Psi(a-0), \\
\hskip 3cm \Psi'(a+0) & = & \Psi'(a-0)+2\,\Omega\Psi(a).
\end{eqnarray} 
From these boundary conditions applied to the wave functions (\ref{eq-psi1}-\ref{eq-psi2}), 
the following homogeneous linear system is obtained
for the unknonws $A$ and $B$:
\begin{equation}
\left(\begin{array}{cc}
\sin ka & \cos ka - e^{iKa} \\
(ke^{iKa}-k\cos ka-2\Omega\sin ka) & (k\sin ka-2\Omega\cos ka)
\end{array}\right)
\left(\begin{array}{c} A \\ B \end{array}\right) = 
\left(\begin{array}{c} 0 \\ 0 \end{array}\right).
\end{equation}
To have a non-trivial solution, the determinant of this $2\times 2$ matrix 
has to be zero. Then, the following quantization relation for $k$ is obtained \cite{flugge1994}:
\begin{equation}
\cos Ka = \cos ka + \frac{\Omega}{k}\sin ka\,.
\label{eq-band}
\end{equation}
The electronic band structure of the one-dimensional crystal is contained in this equation.
When $K$ varies its value in the different Brillouin zones, given by $(m-1)\pi\leq |Ka|\leq m\pi$,
with $m=1$ for the first Brillouin zone, $m=2$ for the second Brillouin zone, etc., 
only certain intervals of $k$ are allowed. These intervals for $k$ are the energy bands of the electronic
system. The positive and negative parts of these intervals correspond with the positive and negative parts 
of the Brillouin zones, respectively.
In the limit $\Omega a=0$, the free electron problem is recovered, then the solutions are
the plane waves with $k=K$. In the limit $\Omega a=\infty$, we have the square well problem,
then the wave number of the eigenstates verify $\sin ka=0$. For an intermediate case, $0<\Omega a<\infty$,
Eq. (\ref{eq-band}) has to be solved. Concretely, for the particular value $\Omega a=4$, that has also been used
in Ref. \cite{flugge1994}, the lower energy bands obtained in this system for $k>0$ are:
\begin{eqnarray}
\hskip 3cm 2.154 \leq ka \leq \;\;\pi & & \hskip 1cm\hbox{(1st band)}, \nonumber\\
\hskip 3cm 4.578 \leq ka \leq 2\pi & & \hskip 1cm\hbox{(2nd band)}, \label{eq-bandas}\\
\hskip 3cm 7.287 \leq ka \leq 3\pi & & \hskip 1cm\hbox{(3rd band)}, \nonumber\\
\hskip 3cm 10.174 \leq ka \leq 4\pi & & \hskip 1cm\hbox{(4th band)}. \nonumber
\end{eqnarray}
The bands are symmetrically found for $k<0$.
Observe that, to finally get the wave function of the electronic states, we additionally
need the normalization condition to completely determine $\Psi(x)$, except a global phase factor.
For our calculations, by taking $\Omega a=4$, we will perform this normalization in the unit cell $[0,a]$.

The basic ingredient to calculate the statistical entropic measures in which we are interested
is the probability density of the electronic states. 
This is given by $\rho(x)=|\Psi(x)|^2$. From this density, 
we proceed to compute the statistical complexity and the Fisher-Shannon information.
Notice that the wave function $\Psi(x)$ for a given $k$ is transformed in $-\Psi(x)$ for $-k$,
therefore all the magnitudes depending on the density are the same in both cases, and then
we reduce our study to the positive part ($k>0$) of the electronic bands. 

Let us recall the definition of the statistical complexity $C$ \cite{lopez1995,lopez2002},
the so-called $LMC$ complexity, that is defined as
\begin{equation}
C = H\cdot D\;,
\end{equation}
where $H$ represents the information content of the system and $D$ gives an idea
of how much sharp is its spatial distribution. 
As a quantifier of $H$, we take a version used in Ref. \cite{lopez2002}. 
This is the simple exponential Shannon entropy \cite{dembo1991},
that has the form, 
\begin{equation}
H = e^{S}\;,
\end{equation}
where $S$ is the Shannon information entropy \cite{shannon1948},
\begin{equation}
S = -\int_0^a \rho(x)\;\log \rho(x)\; dx \;.
\label{eq1-S}
\end{equation}
For the disequilibrium we take the form originally introduced in 
Refs. \cite{lopez1995,lopez2002}, that is,
\begin{equation}
D = \int_0^a \rho^2(x)\; dx\;.
\label{eq2-D} 
\end{equation}

\begin{figure} 
\centerline{\includegraphics[width=12cm]{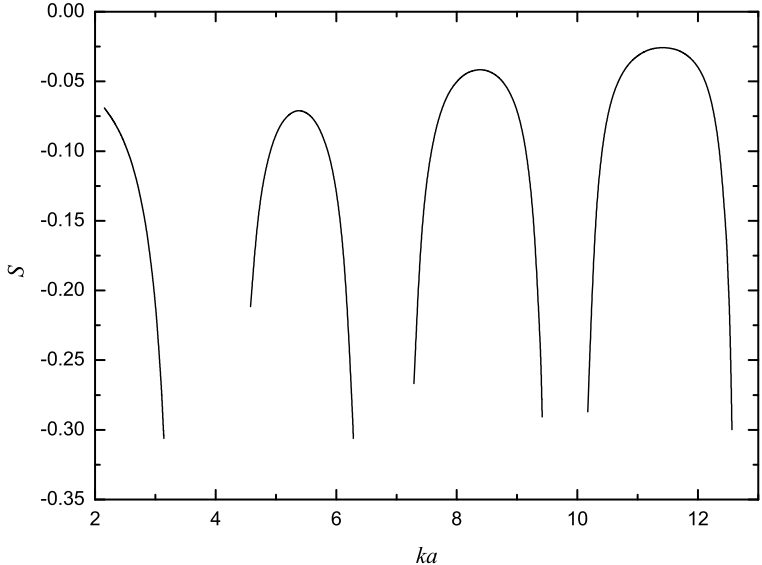}}  
\caption{Shannon entropy, $S$, vs. the adimensional wave number, $ka$, for $k>0$.
Only the four lower electronic bands given in expression (\ref{eq-bandas}) are shown.}  
\label{fig1}  
\end{figure}  

\begin{figure} 
\centerline{\includegraphics[width=12cm]{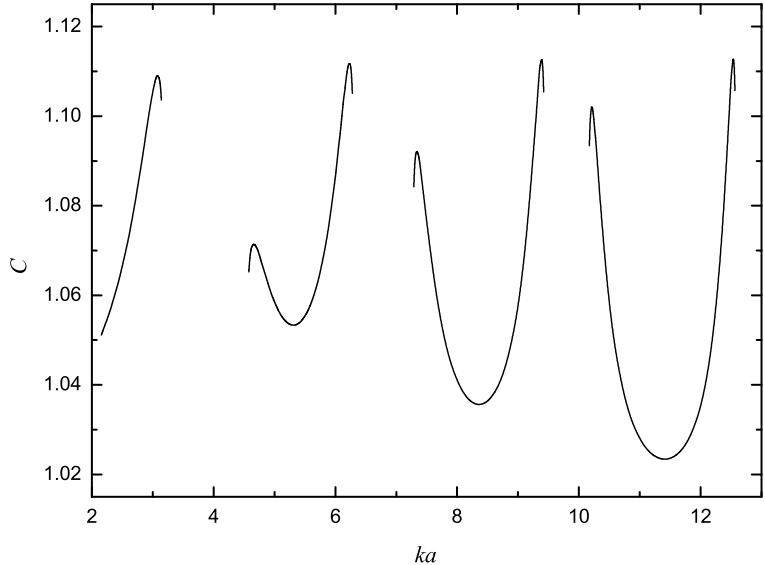}}  
\caption{Statistical complexity, $C$, vs. the adimensional wave number, $ka$, for $k>0$.
Only the four lower electronic bands given in expression (\ref{eq-bandas}) are shown.}  
\label{fig2}  
\end{figure}

The entropy, $S$, and the statistical complexity, $C$, for the lower electronic bands of 
the present one-dimensional crystalline solid are given in Fig. \ref{fig1} and Fig. \ref{fig2}, 
respectively. When this hypothetical solid is in a situation of high conductivity, i.e. when
it contains a band that is partially filled and partially empty, it can be observed in the figures
that the more energetic electrons attain the highest entropy and the lowest complexity in the
vicinity of the half-filled band. This is the point where in general 
the highest conductivity is also attained. Take, for instance, 
the real case of the monovalent metals, that include
the alkali metals (Li, Na, K, Rb, Cs) and the noble metals (Cu, Ag, Au). These metals present 
all the bands completely filled or empty, except an only half-filled conduction band \cite{ashcroft1976}.
Compared with other solids, these metals display a very high conductivity, that in the cases 
of Ag and Cu it is the highest in nature. Then, it is remarkable this coincidence at the point 
of half-filled band where, on the one hand, the entropy and the statistical complexity are 
extrema for this model of solids and, on the other hand, the conductivity reaches its upper values 
for the real cases of monovalent metals. 
 
\begin{figure} 
\centerline{\includegraphics[width=12cm]{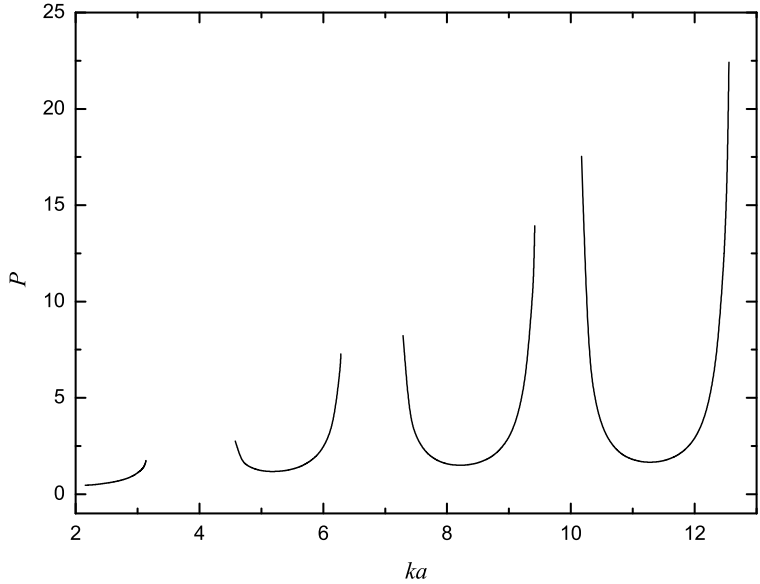}}     
\caption{Fisher-Shannon entropy, $P$, vs. the adimensional wave number, $ka$, for $k>0$.
Only the four lower electronic bands given in expression (\ref{eq-bandas}) are shown.}  
\label{fig3}  
\end{figure}

Now, we check that other statistical entropic measures also display this behavior when the solid
has half-filled bands. Let us take, for instance, the Fisher-Shannon information, $P$, that has been 
applied in different contexts \cite{romera2004,sen2008} for atomic systems. This quantity is given by   
\begin{equation}
P= J\cdot I\,,
\end{equation}
where the first factor
\begin{equation}
J = {1\over 2\pi e}\;e^{2S/3}\;,
\end{equation}
is a version of the exponential Shannon entropy \cite{dembo1991}, 
and the second factor
\begin{equation}
I = \int_0^a{[d\rho(x)/dx]^2\over \rho(x)}\; dx\;,
\end{equation}
is the so-called Fisher information measure \cite{fisher1925}, that quantifies the stiffness 
of the probability density. Observe in Fig. \ref{fig3} the confirmation of the previous results
obtained in Figs. \ref{fig1} and \ref{fig2} for $S$ and $C$, in the sense that the extremal values 
of $P$ for this model of solids are also reached at the half-filling band points.

The former calculations are done orbital by orbital, i.e. thinking that the solid is a set
of individual and independent orbitals, each one identified by its own wave number $k$. 
We can change the point of view of the problem
and to think that the solid stands in some kind of collective state whose probability density 
$\rho_t(x)$ is the normalized sum  of all the allowed electronic densities obtained from the 
orbitals with wave numbers in the interval $[k_{min},k_{max}]$;
$k_{min}$ will be the minimal electronic wave number of the solid, i.e. the lowest $k$ obtained 
in the first band, and $k_{max}$ will be the upper $k$ corresponding to the most energetic electron
of the solid. The expression for $\rho_t(x)$ is
\begin{equation}
\rho_t(x)\,=\,\left.\int_{k_{min}}^{k_{max}}\,\rho_k(x)\,dk \middle / \int_{k_{min}}^{k_{max}}\,dk\right.\,.
\end{equation}
Observe that $\rho_t(x)$ is normalized in the interval $[0,a]$, $\int_0^a\rho_t(x)dx=1$, and
that in the present model of solid $k_{min}=2.154$ as given in formulas (\ref{eq-bandas}).

\begin{figure}  
\centerline{\includegraphics[width=12cm]{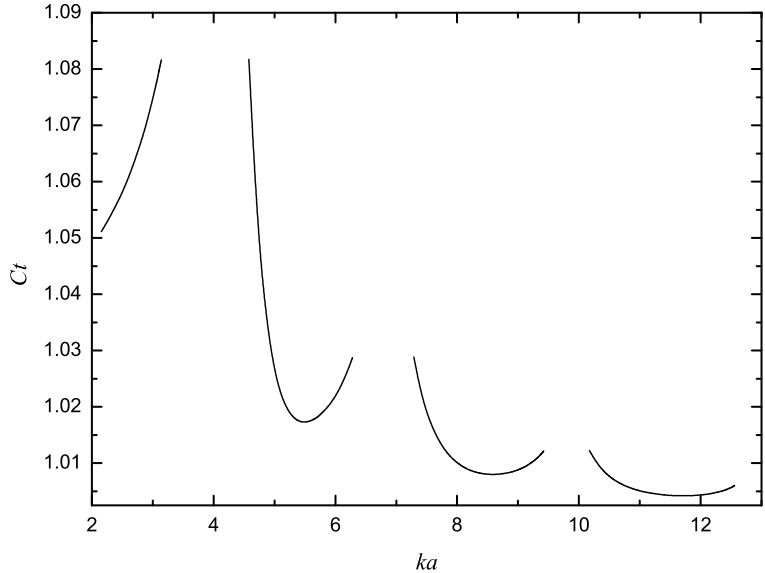}}     
\caption{Accumulated statistical complexity, $C_t$, vs. the adimensional wave number, $ka$, for $k>0$.
Only the four lower electronic bands given in expression (\ref{eq-bandas}) are shown.}  
\label{fig4}  
\end{figure}

The calculation of the statistical complexity $C_t$ for this $\rho_t(x)$ is presented in Fig. \ref{fig4}.
In this case, the minimal values of $C_t$ are also located in the vicinity of the half-filled electronic
bands such as the behavior of $C$ for the individual orbitals shown in Fig. \ref{fig2}.
In the hypothetical limit case of a solid where $k_{max}a\gg 1$, let us remark
that the density $\rho_t(x)$ will tend to the uniform density and then the lowest value 
of complexity, $C_t=1$, can be reached, as it can be seen in Fig. \ref{fig4}.  

In summary, this work puts in evidence that certain conformational properties
of many-body systems are reflected by the behavior of the statistical complexity $C$
and the Fisher-Shannon information $P$. In the present study, the electronic band structure
of a model of solids has been unfolded and the measurement of these magnitudes for such a model
has been achieved. It is remarkable the fact that the extremal values of $C$ and $P$ are attained
on the configurations with half-filled bands, which is also the electronic band configuration 
displayed by the solids with the highest conductivity, let us say the monovalent metals. 
Therefore, the calculation of these statistical indicators for a model of solids 
has unveiled certain physical properties of these systems, in the same way that
these entropic measures also reveal some conformational aspects of other 
quantum many-body systems.

\section*{Acknowledgements}
 The authors acknowledge some financial support from  the Spanish project
 DGICYT-FIS2009-13364-C02-01. J.S. also thanks to the Consejer\'ia de 
 Econom\'ia, Comercio e Innovaci\'on of the Junta de Extremadura (Spain) for 
 financial support, Project Ref. GRU09011.

\end{document}